# Seaborg's Plutonium?

Eric B. Norman, Keenan J. Thomas, Kristina Telhami*

Department of Nuclear Engineering

University of California

Berkeley, CA 94720

## Abstract

Passive x-ray and gamma–ray analysis was performed on UC Berkeley's EH&S Sample S338. The object was found to contain $^{239}$Pu. No other radioactive isotopes were observed. The mass of $^{239}$Pu contained in this object was determined to be 2.0 ± 0.3 µg. These observations are consistent with the identification of this object as containing the 2.77-µg $PuO_2$ sample produced in 1942 and described by Glenn Seaborg and his collaborators as the first sample of $^{239}$Pu that was large enough to be weighed.

* Summer undergraduate research assistant from San Diego State University

In early 1941, in Room 307 Gilman Hall on the University of California's Berkeley campus, Glenn Seaborg and his collaborators Arthur Wahl and Joseph Kennedy chemically separated element 94 from uranium that had been bombarded with 16-MeV deuterons at Berkeley's 60" cyclotron.[1,2] Not long after, Seaborg's group proposed the name plutonium for this new element. For the next year and a half, Seaborg's group performed numerous studies of both the nuclear and chemical properties of plutonium using only trace amounts of this new material. In order to study plutonium in its pure form, an effort was therefore made to produce a "macroscopic" amount of plutonium by irradiating many kilograms of natural uranium with neutrons produced by bombarding beryllium with accelerated deuterons. Cyclotrons at both Washington University and Berkeley were utilized for this production. After chemically extracting the plutonium from the uranium and fission products, on September 10, 1942, at the Metallurgical Laboratory of the University of Chicago, Boris Cunningham and Lewis Werner succeeded in producing the first sample of pure $PuO_2$ that was large enough to be weighed on a newly developed microbalance.[2,3] The mass of the $PuO_2$ was determined to be 2.77 µg. This sample was preserved by sealing the platinum boat and oxide deposit inside a glass tube.[2]

The subsequent history of this sample is not known, but as reported by Richard Strickert[4], it was on display for a number of years at the Lawrence Hall of Science in Berkeley, CA. In 2007, the Lawrence Hall of Science requested that the sample be removed by UC Berkeley's Environment, Health & Safety department to make way for more interactive exhibits. The small plastic box that EH&S removed from the Lawrence Hall had a label on it stating "First sample of Pu weighed. 2.7 µg." The sample was placed into safe and secure storage at the Hazardous Material Facility on the University of Berkeley's campus and was assigned EH&S sample number S338. In 2008, during a routine inventory, UC Berkeley Health Physicist, Phil Broughton, recognized the significance of the sample and contacted several museums to see if they were interested in having it on display. The Smithsonian expressed concern that proof that the sample was what was described on the box was required before they would consider the sample for display. The paper trail documenting this sample's history had been lost and so the question was what could be done to establish its authenticity as Seaborg's plutonium?

In June of 2014, the Nuclear Engineering Department at UC Berkeley became aware of the sample in storage and requested the opportunity to study the sample to determine its origins. In order to preserve the possible historical significance of the sample, it was quickly decided that the box should not be opened and that only non-destructive testing should be done on it. After examining and taking photographs of the sample in its plastic box (shown below in Fig. 1), it was placed approximately 6.3 mm away from the front face of a 36-mm diameter by 13-mm thick planar germanium detector. This detector is equipped with a thin Be window allowing detection of low-energy gamma rays and x rays. The detector was shielded with 1.27 cm of copper and 5 to 10 cm of lead. The box was counted for 21.2 hours. The box was then removed and a background spectrum was collected for 21.3 hours.

Figures 2, 3, and 4 illustrate the relevant portions of the background-subtracted spectrum we obtained from sample S338. All energies are given in keV. All of the gamma rays we observed can be attributed to the decay of $^{239}$Pu (Ref. 5). In order to search for higher energy gamma rays from other radionuclides, the box containing the sample was placed up against the front face of a shielded 85% relative efficiency coaxial Ge detector and counted for approximately 24 hours. No evidence of any additional gamma rays up to 3 MeV was observed. From Figure 3 it can be seen that we observed no evidence of the 59.4 keV gamma ray produced by the decay of $^{241}$Am. This is significant because spent nuclear fuel discharged from power reactors typically contains plutonium with $^{241}$Pu/$^{239}$Pu isotopic ratios in the range of 0.045 - 0.23 (Ref. 6). Even weapons grade plutonium contains $^{241}$Pu/$^{239}$Pu ~ 0.005 (Ref. 7).

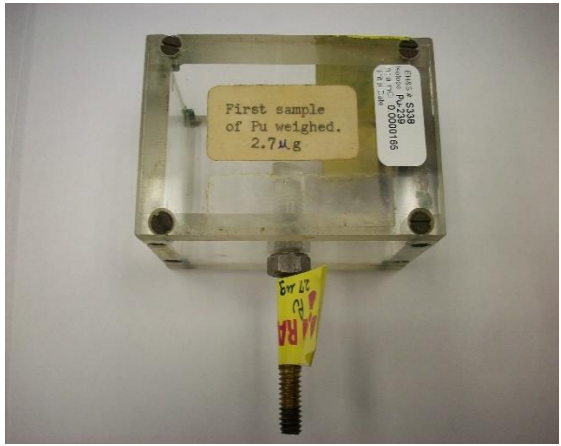

(a)

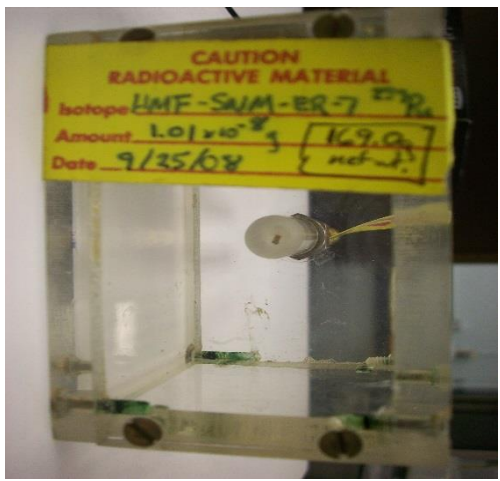

(b)

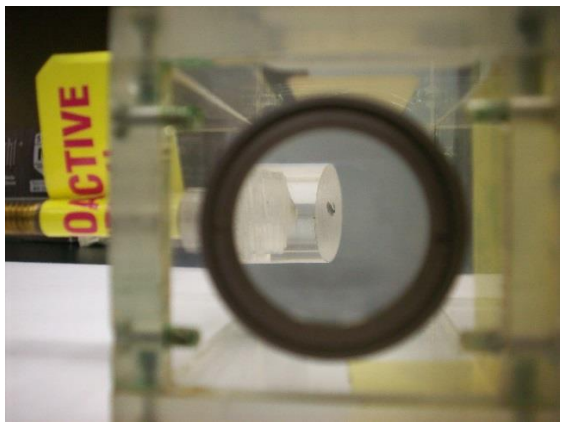

(c)

Fig. 1. (a) Outside of sample box with labels; (b) head-on view showing plastic rod with sample attached; (c) side view showing sample attached to plastic rod.

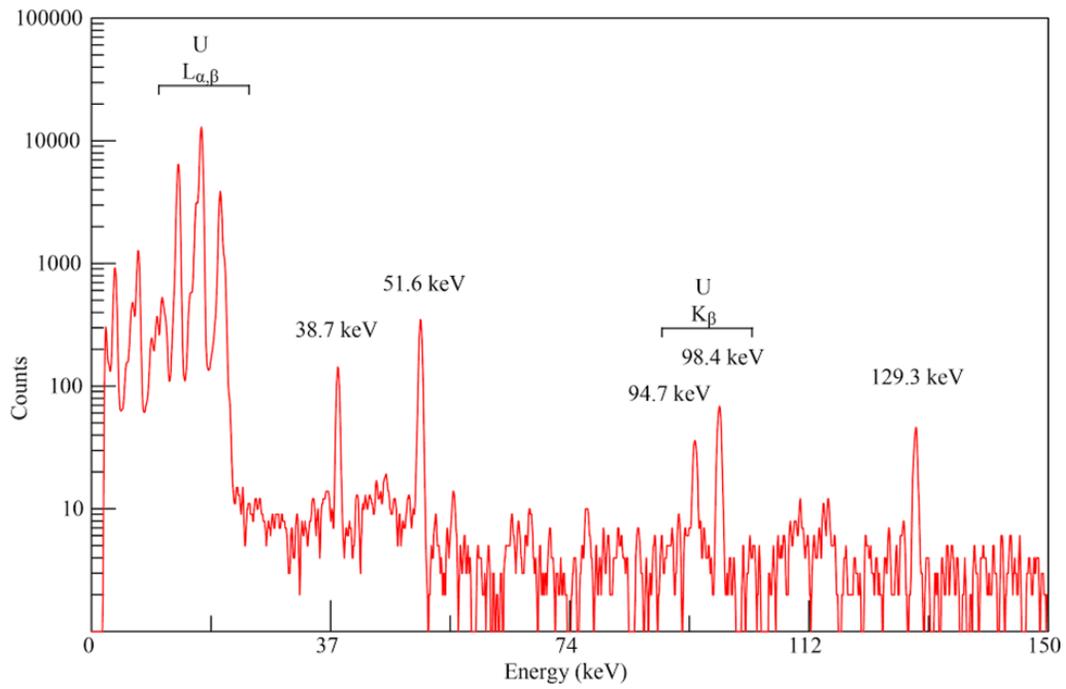

Fig. 2. Background subtracted spectrum observed from Sample S338. All of the labelled peaks are x-rays and gamma rays attributable to the decay of $^{239}$Pu.

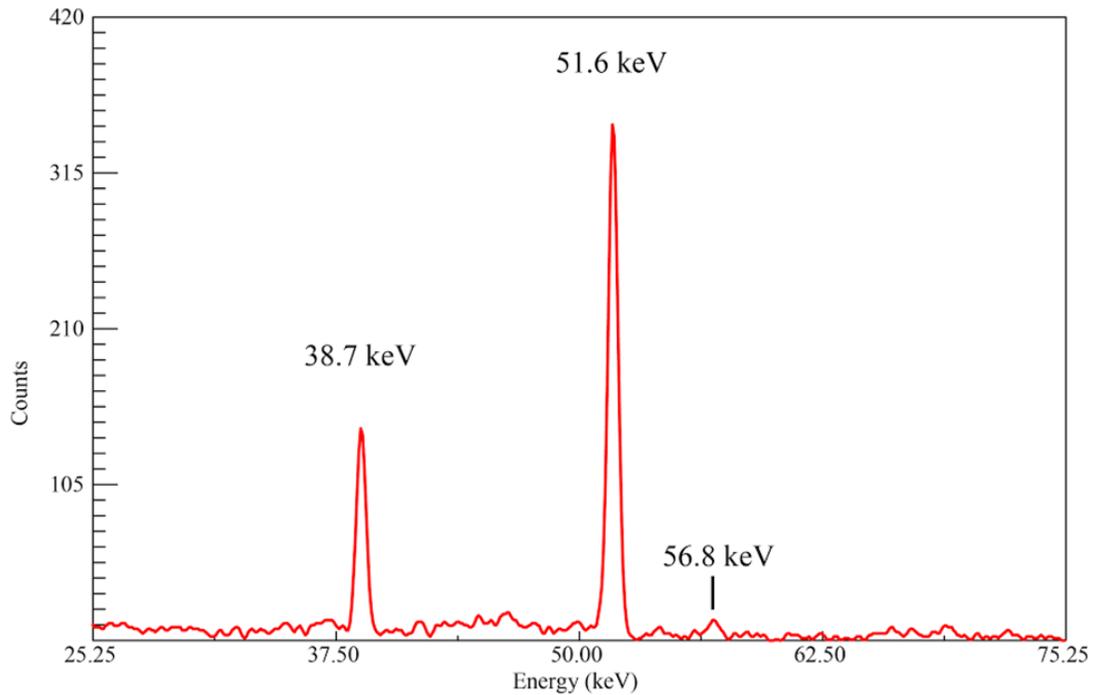

Fig. 3. Expanded region of the spectrum observed around 50 keV. All of the labelled peaks are produced by the decay of $^{239}$Pu. No evidence of the 59-keV gamma ray produced by the decay of $^{241}$Am was observed.

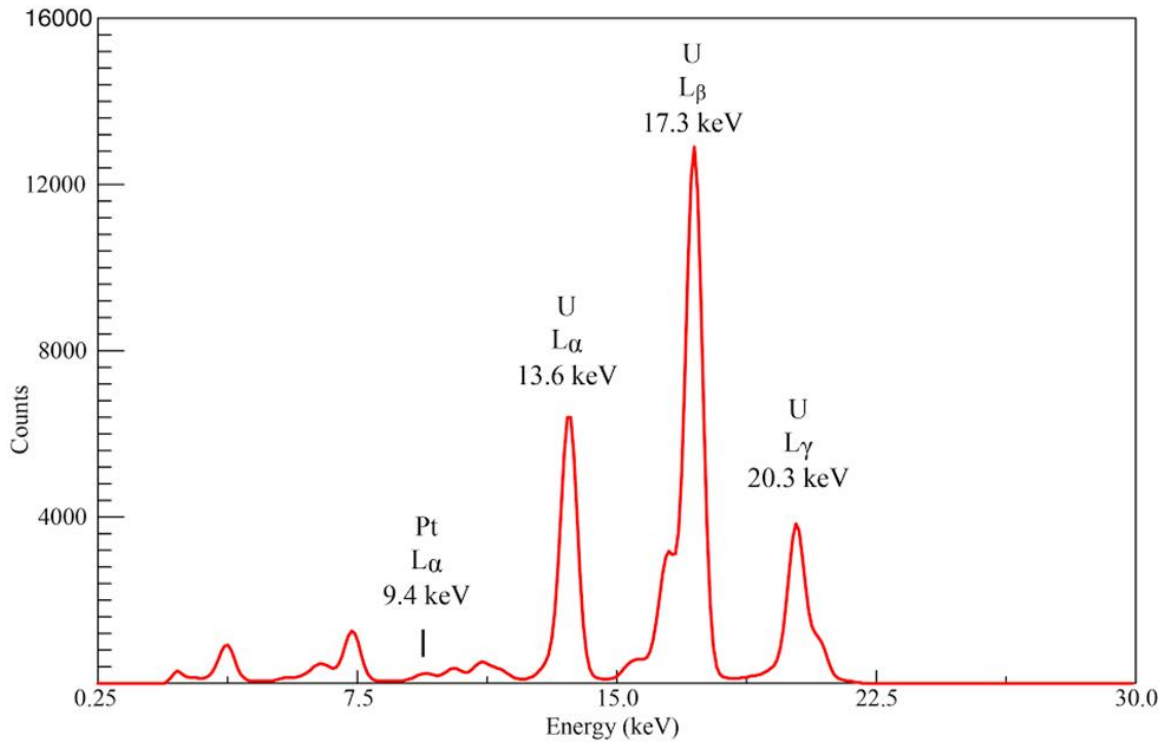

Fig. 4. Low-energy portion of the spectrum showing uranium L x-rays produced by the decay of $^{239}$Pu. The small peak at 9.4 keV is consistent with being an L$_\alpha$ x-ray of platinum. The peaks at lower energies are likely to be Ge escape peaks produced by the higher energy x-rays.

The half-life of $^{241}$Pu is 14.4 years and its decay product is $^{241}$Am. Thus, over time, the abundance of $^{241}$Am in plutonium grows. From our data, we can set an upper limit on the isotopic abundance ratio of $^{241}$Am/$^{239}$Pu in sample S338 to be $< 2.3 \times 10^{-7}$. Thus, the plutonium contained in sample S338 was most likely not produced in a reactor, but in a low neutron fluence environment. This is consistent with Seaborg's description of producing the plutonium by irradiating many kilograms of natural uranium with neutrons produced by deuteron bombardments of beryllium (Ref. 3). In order to determine the mass of $^{239}$Pu contained in this sample, we measured the efficiency of our detector using calibrated sources of $^{57}$Co, $^{137}$Cs, and $^{241}$Am obtained from Isotope Products Laboratories. These sources provide x-ray and gamma-ray lines at 26, 32, 36, 59, 122, and 136 keV, thus covering the energy range of interest. All necessary nuclear data was taken from Ref. 5. We carefully measured the distance from the sample attached to the end of the plastic rod in S338 to the font face of our detector and then placed our calibration sources at this same distance. Gamma-rays emitted from the S338 sample had to pass through the 0.63-cm thick wall of the plastic box in which it is contained. In order to account for the attenuation this produced, we placed a 0.63-cm thick block of polyethylene between our sources and the detector. We extracted the peak areas of the 38.7, 51.6, and 129.3-keV lines from the spectrum shown in Figure 2 and then determined the sample mass from each line. The results were then averaged to establish the mass of $^{239}$Pu contained in sample S338 to be $2.0 \pm 0.3$ µg. Seaborg and Cunningham and Werner stated that the first weighed sample contained 2.77 µg of $PuO_2$ with no uncertainty given. This would imply a $^{239}$Pu mass of 2.44 µg. These authors also stated that the first weighed sample of plutonium was evaporated and then converted to an oxide on a platinum boat (Ref. 2,3). The low energy portion of our spectrum shown in Figure 4 contains a peak at 9.4 keV that is consistent with being the $L_\alpha$ x-ray from platinum. Other still lower energy peaks are most likely Ge K x-ray escape peaks produced by interactions of higher energy x-rays near the surface of our detector.

**Conclusions**

Passive x-ray and gamma–ray analysis was performed on UC Berkeley's EH&S Sample S338. The object was found to contain $^{239}$Pu. No other radioactive isotopes were observed. The mass of $^{239}$Pu contained in this object was determined to be $2.0 \pm 0.3$ µg. Evidence of Pt L x-rays was also seen. While not 100% conclusive, these observations are consistent with the identification of this object being the 2.77-µg $PuO_2$ sample described by Glenn Seaborg and his collaborators as the first sample of $^{239}$Pu that was large enough to be weighed. More detailed measurements such as mass spectrometry or chemical testing could be performed but that would require opening the plastic box and sacrificing some of the plutonium-containing material. It is hoped that in the near future a commemorative display will be set up on the campus or in Seaborg's old office in Gilman Hall where the discovery of plutonium was made. This sample would be a fitting item to showcase in such a setting.

**Acknowledgements**

This worked was supported in part by the U. S. Department of Energy National Nuclear Security Administration under Award Number DE-NA0000979.